\documentclass{JACoW-GSI-2014}
\usepackage{graphicx}
\usepackage[utf8]{inputenc}
\usepackage{amssymb}
\begin{document}
\title{GPU Programming - Speeding Up the 3D Surface Generator VESTA}

\author{B. R. Schlei\thanks{b.schlei@gsi.de}}
\affil{GSI, Darmstadt, Germany}

\maketitle

\begin{abstract}
The novel ``Volume-Enclosing Surface exTraction Algorith'' (VESTA)
generates triangular isosurfaces from computed tomography volumetric images
and/or three-dimensional ($3$D) simulation data.
Here, we present various benchmarks for GPU-based code implementations of both
VESTA and the current state-of-the-art Marching Cubes Algorithm (MCA).
One major result of this study is that VESTA runs significantly
faster than the MCA.
\end{abstract}

\section{Introduction}

NVIDIAs toolkits (\textit{cf., e.g.,} Ref.~\cite{NVIDIA}) for the development 
of CUDA\circledR-based software contain, among many other things, example code
for an extended version~\cite{BOUR94} of the original MCA~\cite{LORE87}.
Here, we compare the performance of this code with our CUDA\circledR- and 
ANSI-C-based implementation of VESTA~\cite{BRS12} on a Linux-based
(\textit{i.e.,} openSUSE $13.1$) PC with a GeForce GTX $750$ Ti graphics card.

In particular, the times that we have measured (\textit{cf.,} Table~1) are
averages over $1000$ runs each.
The measurements start after the data sets have been loaded into texture 
memory, and they stop after all point coordinates and triplets of point IDs
(\textit{i.e.,} triangles) have been computed on the GPU.

\begin{center}
\begin{table}[hb]
\small  
\begin{tabular}{rccc}
\hline
\textbf{Technique}
&Extended MCA
&\multicolumn{2}{c}{Marching VESTA}\\
\textbf{Mode}
&\textbf{DCED / L}
&\textbf{DCED / L}
&\textbf{Mixed / H}\\
\hline
\textbf{(a)$\:\:$ Points}
&$      19,218$
&$      12,814$
&$\:    15,292$\\
\textbf{Triangles}
&$        6406$
&$        6406$
&$\:    11,362$\\
\textbf{Time (ms)}
&{\boldmath$  1.43(5)$}
&{\boldmath$  1.28(5)$}
&$\:  1.37(4)$\\
\hline
\textbf{(b)$\:\:$ Points}
&$   6,128,724$
&$   4,085,840$
&$\: 4,852,644$\\
\textbf{Triangles}
&$   2,042,908$
&$   2,042,908$
&$\: 3,576,516$\\
\textbf{Time (ms)}
&{\boldmath$ 98.5(1)$}
&{\boldmath$ 71.3(1)$}
&$\: 75.9(4)$\\
\hline
\textbf{(c)$\:\:$ Points}
&$   5,566,998$
&$   3,699,086$
&$\: 4,346,120$\\
\textbf{Triangles}
&$   1,855,666$
&$   1,855,666$
&$\: 3,147,604$\\
\textbf{Time (ms)}
&{\boldmath$ 23.0(1)$}
&{\boldmath$ 18.7(1)$}
&$\: 22.4(1)$\\
\hline
\textbf{(d)$\:\:$ Points}
&$   33,240$
&$   22,208$
&$\: 25,894$\\
\textbf{Triangles}
&$   11,080$
&$   11,080$
&$\: 18,350$\\
\textbf{Time (ms)}
&{\boldmath$ 0.82(4)$}
&{\boldmath$ 0.63(4)$}
&$\: 0.74(4)$\\
\hline
\textbf{(e)$\:\:$ Points}
&$   13,859,304$
&$    9,267,824$
&$\: 11,178,649$\\
\textbf{Triangles}
&$   4,619,768$
&$   4,619,768$
&$\: 8,441,610$\\
\textbf{Time (ms)}
&{\boldmath$ 111.2(1)$}
&{\boldmath$  84.4(1)$}
&$\: 94.6(6)$\\
\hline
\end{tabular}
\caption{
Benchmarks for various processed tomographic data sets: 
for (a) -- (c), \textit{cf.,} Ref.~\cite{BRS12} and Ref.s therein, 
(d) Bucky.raw data is a portion of~\cite{NVIDIA}, 
and (e) Happy Buddha VRI file~\cite{HAPPY}.
For the selected isovalues, \textit{cf.,} Fig.~1.
}
\label{tab01}
\end{table}
\end{center}

\vspace{-1.3cm} 

\begin{figure}[tp]
\centering
\includegraphics*[width=82mm]{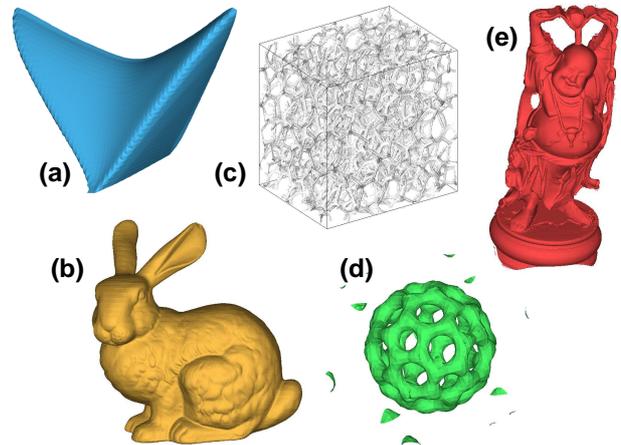}
\caption{VESTA high resolution ``mixed'' mode (Mixed/ H) 
isosurface renderings, where the isovalues equal to (a) $139$, (b) $150$, 
(c) $135$, (d) $128$, and (e) $150$, respectively.}
\label{fig01}
\end{figure}

\section{Results}

For the here considered data sets~\cite{NVIDIA,BRS12,HAPPY}, the extended MCA 
is about (a) $12\%$, (b) $38\%$, (c) $23\%$, (d) $30\%$, and (e) $32\%$,
\textit{slower} than the marching variant of VESTA~\cite{BRS12}, 
when the latter is executed in its low resolution ``disconnect'' mode (DCED/L).
Furthermore, VESTA is also faster even if higher resolution isosurfaces
are computed (\textit{cf.,} Fig.~1), which have about twice the number of
triangles (\textit{cf.,} Table~1).

Note that the current code implementation of VESTA does \textit{not} yet use
parallel streaming, \textit{nor} it does call device kernels from within
kernels.
As a consequence, further GPU-based code optimisations may result in an even
faster VESTA code.

\end{document}